\documentclass[aps,prb,showpacs,twocolumn,groupedaddress]{revtex4-1}
\usepackage{times,xspace}
\usepackage{amsfonts,amsmath}
\usepackage{amsfonts}
\usepackage{bm}
\usepackage{amssymb}
\usepackage[final]{graphicx}
\usepackage{color}

\begin{document}

\title{Self Energy and Fluctuation Spectra in Cuprates:
Comparing Optical and Photoemission Results}
\author{R.S. Markiewicz$^1$, Tanmoy Das$^{1,2}$, and A. Bansil$^1$}
\affiliation{$^1$Physics Department, Northeastern University, Boston, MA, 02115, USA.\\$^2$Theoretical Division, Los Alamos National Laboratory, Los Alamos, NM, 87545, USA.}
\date{\today}
\begin{abstract}
We compare efforts to extract self energies and fluctuation spectra of the cuprates using optical and photoemission techniques.  The fluctuations have contributions from both the coherent and incoherent parts of the band, which are spread over the full bare bandwidth of ~$>2$~eV.
Many experimental studies concentrate on the coherent part of the band and hence miss higher energy fluctuations. Our study
establishes the universal presence of high energy bosonic fluctuations across various
spectroscopies as a key ingredient in the high temperature superconducting cuprates.
\end{abstract}
\pacs{78.20.-e,78.15.+e,74.25.Gz,71.10.-w} \maketitle

\section{Introduction}

A critical issue in unlocking the mechanism of superconductivity in the
cuprates is determining  the spectrum of bosonic [phononic or electronic] fluctuations that strongly interact with the electrons, and which can drive a variety of instabilities and exotic physics.
In particular, what is the energy scale of the relevant fluctuations: If high-$T_c$
superconductivity is produced by fluctuations at a low energy scale
comparable to the magnetic resonance mode\cite{resy2}, then the bosons
responsible for pairing -- the so-called `glue' -- could be phonons or
magnetic excitations.  If on the other hand higher energy fluctuations
($\sim J$, $U$, or charge transfer scales) play an important role\cite{PWA,glue,MRsc,Emery}, a
novel electronic mechanism would be clearly indicated.

Recently there have been a number of attempts to extract self energies and fluctuation spectra of the cuprates from angle-
resolved photoemission (ARPES), tunneling, and optical spectra.  Most experimental probes find strong coupling to a
low-energy boson\cite{SchaCar,DoHoTV,Casek,NoChu,Marel,GDM} -- often with significant isotope
effect\cite{Seam,Lanz,DanD}.   In addition, ARPES finds a high-energy kink (HEK) suggestive of significant coupling to electronic bosons in the 300-600~meV range.\cite{RonK,Ale,Non,Valla,Feng,Zhou,Kordyuk,Ronning}
These bosons are believed to be predominantly magnetic,\cite{water3,waterfall} with charge bosons at higher energies up to the charge-transfer energy $\sim$2~eV.
In contrast, recent attempts to extract the fluctuation spectra or the `optical glue'\cite{foot1} functions from optical measurements\cite{SchaCar,DoHoTV,NoChu,Marel,GDM} find little evidence for spectral weight above $\sim$300~meV, suggesting that high energy scales are unimportant in the cuprates. However, these analyses are generally restricted to energies below $\sim$1~eV, thereby precluding the possibility of fluctuations at the $\sim$2~eV charge transfer energy scale (the $U$ scale of the one band Hubbard model).\cite{foot4}

Several recent attempts at quasi-first-principles calculations of the optical spectra have found that the cuprate intraband optical spectrum extends up to $\sim$2.5~eV, with a residual charge transfer gap, associated with the incoherent part of the band, persisting well into the overdoped regime\cite{comanac,tanmoyop,DMFT1,DMFT2}.  This suggests that optical glue studies should be extended into the higher energy regime, to provide definitive answers about the role of charge transfer excitations in high-$T_c$ superconductivity.  This is important since several optical studies\cite{Little,GDM} have found evidence that the onset of superconductivity affects spectral weight in an energy range extending beyond 1~eV.

In the present paper, we explore how the results of realistic self energy calculations can be used to guide the analysis of the optical glue.  We explore the role of anisotropy, and clarify just what the glue function actually measures.  We point out a simple correction to commonly used formulas for self energy, which largely eliminates the problem of negative scattering rates.  We find that an important contribution to the high-energy fluctuations has been `hiding in plain sight'.  And we provide an example of an optical glue recovery which includes the high energy contribution, and which bears a striking similarity to the calculated results.  Our study thus resolves a puzzling
discrepancy between optical vs other experiments related to the nature of
bosonic fluctuations, and clearly demonstrates that experimental attempts to extract the optical glue
need to probe a higher energy range to weigh in on the issue of a possible $U$-scale glue
 involved in the mechanism of high-$T_c$ superconductivity.

This paper is organized as follows.  Section IIA shows how including self energy effects in optical calculations can provide insight into attempts to extract fluctuation, or `pairing glue'  functions from optical experiments. Section IIB provides an example of extracting glue functions from real optical data.  Section III describes numerical experiments to test the accuracy of the glue extraction, while Section IV compares the optical self energy to ARPES-derived self energies. Sections V and VI give a discussion and conclusions, respectively.  Details of our self energy calculations are presented in an Appendix.

\section{Calculating the `Glue' Function}
\subsection{General Considerations}

We first briefly comment on how the optical `glue' is measured and what it really represents.  One starts with the optical conductivity $\sigma(\omega)$, which can be calculated following Allen\cite{Alle}.  For a $k$-independent $\Sigma$
\begin{eqnarray}\label{sigma}
\sigma(\omega) = \frac{i\omega_p^2}{4\pi\omega}\int_{-\infty}^{\infty} d
\omega^{\prime}\frac{n_F(\omega^{\prime})-n_F(\omega^{\prime}+\omega)}
{\omega+\Sigma^*(\omega^{\prime})-\Sigma (\omega^{\prime}+\omega)},
\end{eqnarray}
where $n_F$ is the Fermi function.  In optical glue studies, an `optical self energy' function is derived from the experimental
optical conductivity $\sigma$ by assuming it to be of an extended Drude form\cite{hwangnature,hwangprl}
\begin{equation}
\sigma(\omega) =
\frac{i\omega_p^2}{4\pi}\frac{1}{\omega-2\Sigma_{op}(\omega)},
\label{eq:0}
\end{equation}
where $\omega_p=\sqrt{4\pi ne^2/m}$ is the plasma frequency, $n$ the
carrier density, and $e$, $m$ are the electronic charge and mass
respectively.  This can also be written in terms of a frequency dependent
scattering time $\tau(\omega)$ and effective mass $m^*(\omega)$ as
\begin{equation}
2\Sigma_{op}(\omega) = \omega (1-{m^*\over m}) - {i\over\tau}.
\label{eq:0a}
\end{equation}

An important question is, how is $\Sigma_{op}$ related to the self energy function $\Sigma$?
Two approaches are commonly employed, which we call Method I, which simply assumes:
\begin{equation}
\Sigma_{op} =\Sigma,
\label{eq:2f}
\end{equation}
 and Method II:\cite{PBA2,Mars,Shulga,HEscale}
\begin{equation}
\Sigma_{op} =\frac{\int_0^{\omega}\Sigma (\omega')d\omega'}{\omega}.
\label{eq:2g}
\end{equation}
We test these schemes in Section III, by comparing self energies extracted from calculated optical spectra with the input self energies, calculated using a GW model\cite{Hedin} appropriate for the cuprates.  We find similar results for both methods: they can approximately reproduce $\Sigma"$ at energies below $\sim$1~eV, but both methods have problems in the pseudogap regime.

In the GW method, the self energy is calculated as the convolution of the Green's function $G$ and an interaction $W$
which is $U^2$ times a susceptibility, which represents the spectrum of electronic bosonic modes (see the Appendix for details).  For cuprates with $d$-wave pairing, the true pairing glue is given by the $d$-wave average of this $q$-dependent function.  In contrast, the optical spectra are measured at $q=0$, and provide no information on this anisotropy.  Hence optical experiments can only measure an average $\bar W$, which we nevertheless denote as $\alpha^2F$.
Below we show that $\alpha^2F$ approximately represents the $q$-averaged susceptibility
\begin{equation}
\alpha^2F(\omega)= U^2[\bar\chi"_c(\omega)+3\bar\chi"_s(\omega )]/2,
\label{eq:00a}
\end{equation}
 with $U$ the Hubbard $U$, $\chi"$ the imaginary part of $\chi$, $\chi_c$ [$\chi_s$] the charge [spin] susceptibility,
and we neglect the distinction between $\alpha^2F$ and $\alpha_{tr}^2F$, where the latter is a transport Eliashberg function.   Also $\bar\chi_i (\omega )=\int a^2d^2q\chi_i ({\bf q},\omega )/(2\pi)^2$, $i=s,c$.  Within RPA
\begin{equation}
\chi_s =\chi_0/(1-U\chi_0),
\label{eq:00b}
\end{equation}
\begin{equation}
\chi_c = \chi_{0}/\epsilon,
\label{eq:00c}
\end{equation}
where $\chi_{0}$ is the bare susceptibility as calculated in the local-density approximation [LDA], and the dielectric constant can be written as
$\epsilon =\epsilon_0+U\chi_{0}$ with $\epsilon_0\sim 4.8$, a background
dielectric constant.  Thus optical measurements determine the $q$-averaged susceptibility as a function of energy.  This can be compared to susceptibilities measured in other spectroscopies, such as ARPES, to determine the spectra of the bosons which strongly couple to electrons, but cannot directly provide information on how well these bosons contribute to $d$-wave superconductivity.

When the susceptibility is replaced by its $q$-average, the formula for the self energy can be simplified, Appendix A.  At $T=0$, for $\omega >0$, it becomes
\begin{equation}
\Sigma^{\prime\prime}(\omega )=-\int_0^{\omega}\alpha^2F(\Omega
)\tilde N(\omega-\Omega )d\Omega , \label{eq:4}
\end{equation}
where $\tilde N(\omega )=[N(\omega)+N(-\omega)]/2N_{av}$ is the electron-hole averaged density of states (DOS), and $N_{av}$ is the average of the DOS over the energy range of interest, chosen to make $\tilde N$ dimensionless.
Then for Method II
\begin{equation}
\Sigma_{op}^{\prime\prime}(\omega)=\int_0^{\omega}\frac{d\Omega}{\omega}\alpha^2F(\Omega
)n_{eh}(\omega-\Omega),
\label{eq:3A}
\end{equation}
where
\begin{equation}
n_{eh}(\omega)=\int_0^{\omega}d\omega \tilde N(\omega).
\label{eq:4A}
\end{equation}

We note that in many previous inversion schemes the DOS is approximated by a constant.
In this case Eq.~\ref{eq:4} becomes\cite{foot6}
\begin{equation}
\alpha^2F_0(\omega)=-{\partial\Sigma_{op}''(\omega )\over\partial\omega}.
\label{eq:2}
\end{equation}
We use the subscript `0' on $\alpha^2F$ to denote that it is assumes a constant DOS.
According to Eq.~\ref{eq:2} the glue
function can become negative unless $|\Sigma_{op}^{\prime\prime}|$ is a
monotonically increasing function of $\omega$.\cite{foot2} We find that
$\alpha^2F$ and $\alpha^2F_0$ can display very different energy
dependencies as can be seen by comparing Figures~\ref{fig:3}(a) and \ref{fig:3}(b) below.  A similar problem arises with
Method II.\cite{PBA2}
Many groups use a finite-temperature version of this result\cite{GDM,Maksimov}.  From Eq.~\ref{eq:3A}, it follows that $-\partial\Sigma_{op}"/\partial\omega$ must be $>0$, as for Eq.~\ref{eq:2}, and
\begin{equation}
\alpha^2F_1(\omega)=-\frac{\partial}{\partial\omega}\Bigl(\omega^2\alpha^2F_0(\omega)\Bigr),
\label{eq:2A}
\end{equation}
where $\alpha^2F_1$ is the glue function corresponding to Eq.~\ref{eq:3A}.  While this works for phonon contributions to the self energy, this substitution is not appropriate over a 2-3~eV energy range.
We find that $\Sigma"$ cannot be monotonic over the full bandwidth, and that may be why previous analyses often had problems with negative $\alpha^2F$, or why they are restricted to fairly low energies.  In contrast, these problems do not arise with Eqs.~\ref{eq:4},~\ref{eq:3A}.

The question remains, what is the appropriate DOS?  This has two aspects.  First, since the optical self energy $\Sigma"_{op}$ involves the sum of the electron and hole
self energies, one should use the average of the electron and hole DOS, $\tilde N$ in Eq.~\ref{eq:4}.
But which DOS?  In the GW approach, one could consider using either the bare $N_0$ or the fully dressed DOS $N$.  The use of $N$ is prohibitively expensive, since to calculate it requires knowledge of the glue function, which we are trying to extract. On the other hand, $N_0$ can be directly calculated from the LDA dispersion and should work progressively better with overdoping, as correlation effects weaken.  For simplicity, in the present analysis we use $N_0$.  This approximation should have no effect on the qualitative features we are describing.

\subsection{Application to Bi2212}

\begin{figure}[htop]
\hspace{-1cm}
\rotatebox{0}{\scalebox{.6}{\includegraphics{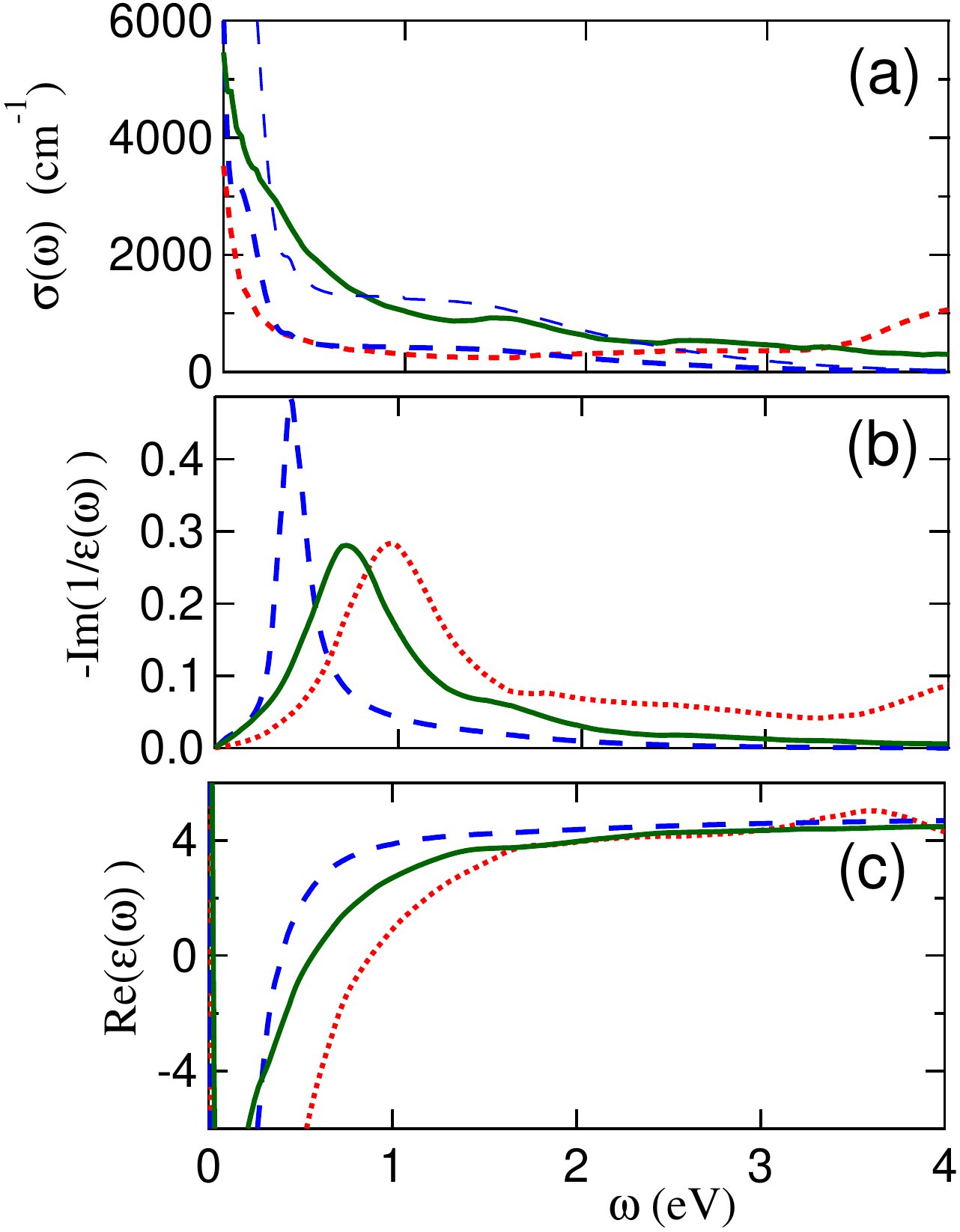}}}
\caption{(color online) (a) Optical conductivity of optimally-doped
Bi2212 [$a$-axis]\cite{QT} (red short-dashed line) and $x=0.15$
LSCO\cite{Uchida} (green solid line) compared with theory for
LSCO\cite{tanmoyop} (blue dashed line; expanded $\times 3$:  thin blue dashed line). (b) Corresponding loss function,
$-Im(1/\epsilon)$.  (c) Dielectric constant $\epsilon$. }
\label{fig:00}
\end{figure}

Figure~\ref{fig:00}(a) compares the measured\cite{QT} optical conductivity of near-optimally-doped  Bi$_{2}$Sr$_{2}$CaCu$_{2}$O$_{8+\delta}$ (Bi2212) (red dotted line)\cite{footC1,Aoki} and
La$_{2-x}$Sr$_x$CuO$_4$ (LSCO)l\cite{Uchida} (green solid line) with calculated\cite{tanmoyop}
(blue dashed line) conductivities for LSCO. The theoretical conductivity consists of a Drude term associated with
the coherent part of the electronic band plus an effective interband term associated with the incoherent spectral weight, a residue of the upper and lower magnetic bands.  While the theory
underestimates the incoherent spectral weight [a known shortcoming of
the QP-GW model]\cite{footy,markiecharge}, it does capture an enhanced conductivity near 1.5~eV,
associated with this residual charge transfer band\cite{tanmoyop}.  
Given the conductivity, the corresponding dielectric constant and inverse dielectric
constants can readily be calculated.  These are shown in Figs.~\ref{fig:00}(b,c) for
LSCO, to be compared to the experimental results for Bi2212.  The good
agreement strongly suggests that all features in the spectrum up to
$\sim$2.5~eV are characteristic features of the cuprate plane, and are
well described by a single band Hubbard model.\cite{footC2}
Note that the electronic susceptibility should be nonzero over the same frequency window as the loss function, so both charge and spin contributions to the glue function should remain finite up to $\sim$2.5~eV, which is approximately the bare [LDA] electronic bandwidth.

\begin{figure}[htop]
\hspace{-1cm}
\rotatebox{0}{\scalebox{.65}{\includegraphics{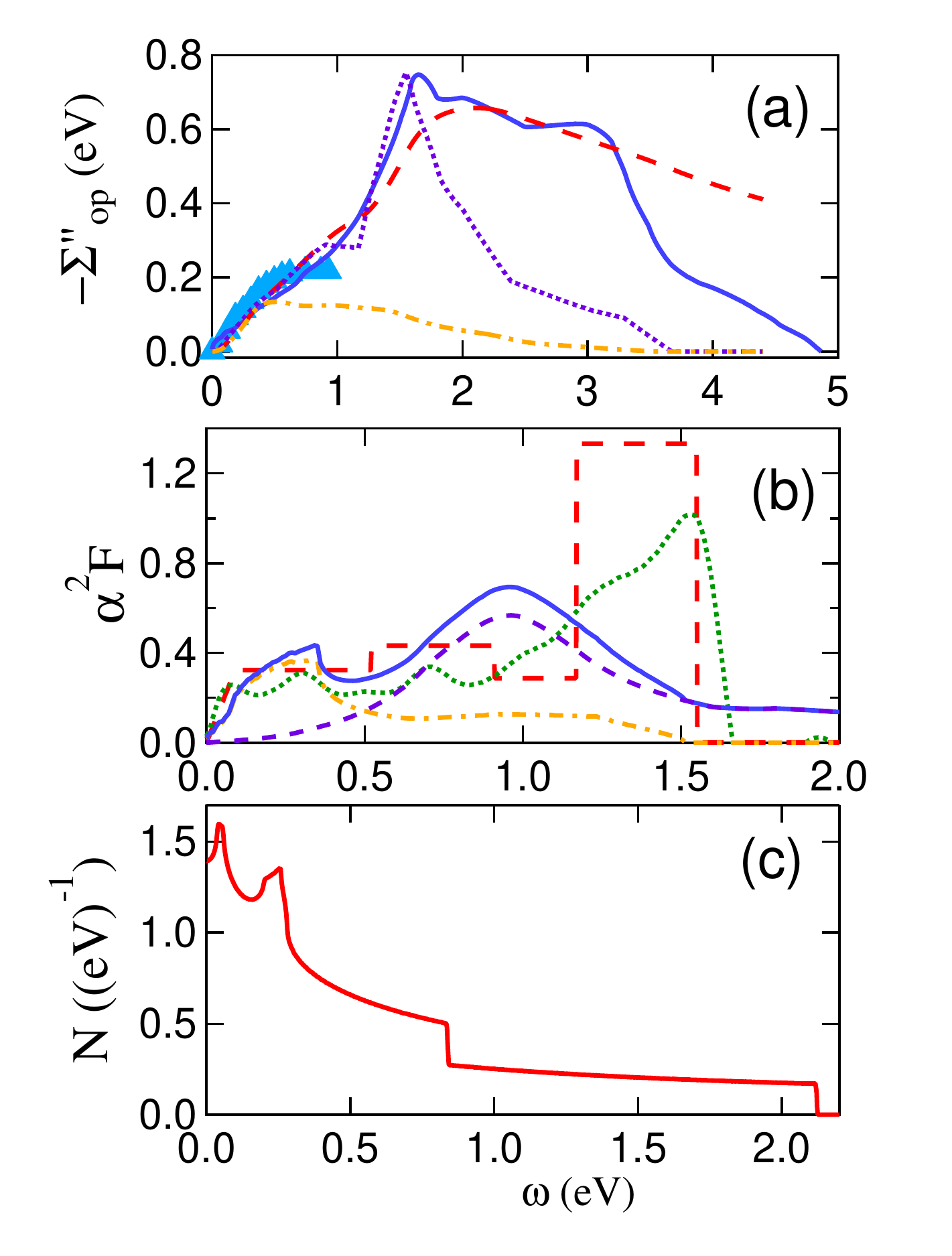}}}
\caption{(color online) (a)
Optical self energy extracted from the experimental data\cite{QT} of Fig.~\ref{fig:00} (dark blue solid line), compared with calculated self energies
using Eq.~\ref{eq:3A} (red dashed line) or Eq.~\ref{eq:4} (violet dotted line) and the corresponding glue function from
frame (b) (red dashed line).  Also shown is the low-energy measured self energy\cite{vdMMZ} (blue triangles) from Fig.~\ref{fig:0} below.
(b) Glue functions used in frame (a).  Green dotted line is glue function calculated from Eq.~\ref{eq:2}. These are compared  with calculated glue functions, including the calculated spin $\alpha^2F_s$  of Fig.~\ref{fig:3}(b) (orange dot-dashed line), the charge glue function, Eq.~\ref{eq:00d}, with $U_{c,eff}=2~eV$ (violet dashed  line), and their sum (blue solid line).
(c) Electron-hole averaged bare DOS for Bi2212.
}
\label{fig:3b}
\end{figure}

Given the optical spectrum, we can extract the glue function over the full bandwidth.
Figure~\ref{fig:3b}(a) plots the optical self energy of Bi2212  (solid blue line) extracted from the data in Fig.~\ref{fig:00}\cite{QT} using Eq.~\ref{eq:0}.  For simplicity, we calculated $\Sigma_{op}$ using the full optical spectrum, but the features above $\sim$2.5~eV are probably due to interband transitions, and hence should be disregarded.
In the low energy limit, the self energy is in reasonable agreement with earlier work\cite{vdMMZ} (blue triangles), which neglected the DOS factor.  Shown also are model self energies calculated from Eq.~\ref{eq:4} or Eq.~\ref{eq:3A}, using the glue function plotted in Fig.~\ref{fig:3b}(b) as a red dashed line and the DOS of Fig.~\ref{fig:3b}(c).  For the present illustrative purposes, we use $T=0$ expressions to analyze the spectrum, even though the experiments were done at room temperature.

In our model calculation, we represent the glue function by a simple histogram, red dashed line in Fig.~2(b).  We find a very good fit to the data using Method II, black dot-dot-dashed line in Fig.~\ref{fig:3b}(a).  Using the same glue function, the self energy calculated by Method I (violet dotted line in Fig.~\ref{fig:3b}(a)) fits the data well below 1.5~eV, but falls off too rapidly at higher energies.  An improved fit would therefore require additional weight in the glue function at even higher energies, which seems less likely.  Remarkably, the green dashed line in Fig.~\ref{fig:3b}(b) shows that the glue function can also be approximately found from Eq.~\ref{eq:2}, if we neglect the negative glue function contributions at higher energy.

According to Eq.~\ref{eq:00a}, the glue function should be calculable in terms of the spin and charge susceptibilities.
In a previous publication\cite{markiecharge} we showed that the charge susceptibility that enters the self-energy should be the same as in the loss function plotted in Fig.~1(b).  However, From Fig.~1(b), it can be seen that the present model does not well reproduce the experimental loss function, perhaps because the Hubbard model does not describe the charge susceptibility of Eq.~\ref{eq:00c} well, and longer range Coulomb interactions need to be included.\cite{footy,water3,MarB2} We can avoid this difficulty by directly comparing two experimental measures of the loss function, taking the charge contribution to the glue function as
\begin{equation}
\alpha^2F_{c}(\omega)=-\frac{U_{c,eff}}{2}Im(\frac{1}{\epsilon (\omega)}),
\label{eq:00d}
\end{equation}
where $U_{c,eff}$  is a phenomenological charge vertex, and $\epsilon$ is the measured dielectric constant.
In Fig.~\ref{fig:3b}(d), we compare the extracted glue function with the calculated spin susceptibility
of Fig.~\ref{fig:3}(b) (orange dot-dashed line) [the corresponding self energy is plotted in  Fig.~\ref{fig:3b}(a) (orange dot-dashed line)].  There is satisfactory agreement up to 0.5~eV, but the experimental glue function reveals excess weight in the charge-transfer regime above 1~eV.  There should be an extra contribution due to charge fluctuations, which we plot in Fig.~\ref{fig:3b}(b),  using the experimental loss function from Fig.~\ref{fig:00}(b) in Eq.~\ref{eq:00d}.
Figure~\ref{fig:3b}(b) shows that the combination of spin and charge glue functions qualitatively reproduces the glue function extracted directly from the optical spectrum, including a significant contribution near 1~eV.  Differences above 1~eV may be due to limitations in extracting the true $\Sigma$ from $\Sigma_{op}$, as discussed in the following section.  We note that the contribution of the loss function to the optical glue has not been previously recognized.

\section{Testing Optical Inversion Schemes}

\subsection{Connecting $\Sigma$ and $\alpha^2F$}

Having a realistic scheme for calculating optical spectra allows us to test how well glue functions can be extracted, by inverting calculated data and comparing the inverted with the input susceptibilities.  Here we briefly describe the results of several tests we have carried out.  We first note that while the $q$-dependence of $\chi"$ is important, the self energy is a convolution over $\chi$ and the Green's function, and we find that its momentum-dependence is relatively weak in the overdoped regime.  This is important, since when the self energy has a significant momentum dependence, the Kubo formula for conductivity includes significant vertex corrections\cite{ChaMu}, which to our knowledge have not been included in any inversion scheme. Fortunately, the weak momentum dependence of the self energy that we find suggests that a self energy extracted from optical studies could still be fairly representative.

\begin{figure}[htop]
\hspace{0cm}
\rotatebox{0}{\scalebox{.37}{\includegraphics{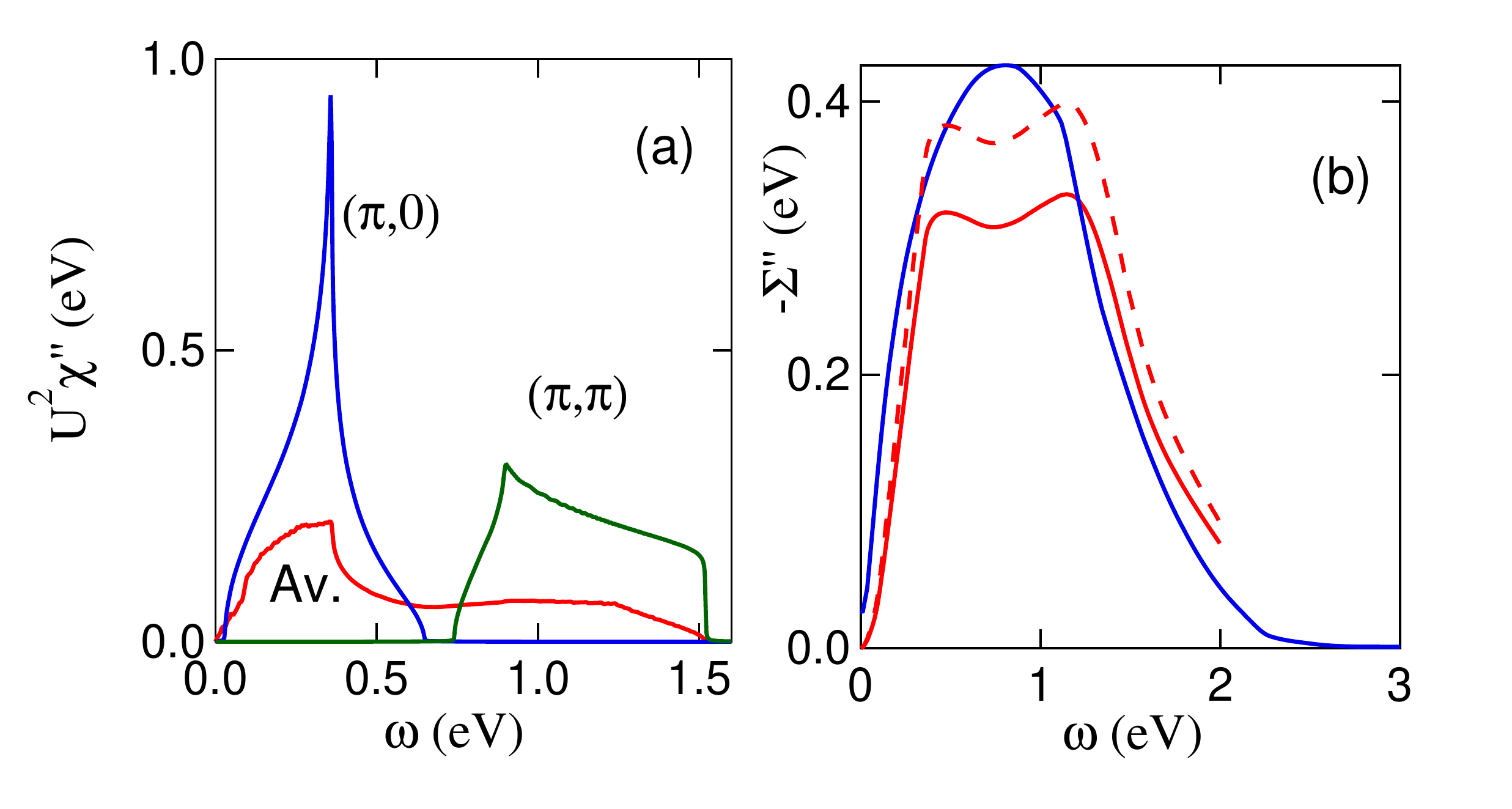}}}
\caption{(color online) (a) Imaginary susceptibility
$\chi_s^{\prime\prime}$ at $T=0$ plotted along two directions in
$k$-space, $(\pi ,0)$ (blue line) and $(\pi ,\pi )$ (green line),
along with the value averaged over all $k$ (red line). (b) Resulting imaginary
self energy $\Sigma^{\prime\prime}$, comparing the exact
anisotropic calculation (blue line) with the present isotropic
approximation (red line), and a scaled isotropic calculation
multiplied by 1.2 (dashed red line).
} \label{fig:1}
\end{figure}

Since optical techniques can only find a momentum-averaged glue function, in Fig.~\ref{fig:1} we address the issue of the
extent to which the average glue function can reproduce the self energy.\cite{foot3,MGu2,foot9}
We denote this momentum-averaged glue function as $\alpha^2F (\omega)$.  To avoid tensor complications we limit
our analysis to the overdoped normal state spectra and analyze only the RPA magnetic susceptibility
$\chi_s =\chi_0/(1-U\chi_0)$, with bare susceptibility $\chi_0$ and $U$
being the Hubbard $U$. In Fig.~\ref{fig:1}(a), the average $\bar\chi_s^{\prime\prime}$
is compared to individual $\chi_s^{\prime\prime}$ peaks, demonstrating that
there is significant anisotropy in $\chi_s^{\prime\prime}$.
The weight in
$\bar\chi_s^{\prime\prime}$ extends to energies $\sim$1.5~eV, in
good agreement with early optical determinations\cite{Little}.
Figure~\ref{fig:1}(b) compares the calculated $\Sigma^{\prime\prime}$ using
either the correct expression, Eq.~\ref{eq:3b} below (blue line), or the
angle averaged Eq.~\ref{eq:4} value (red line).  It can be seen
that averaging before integrating underestimates the magnitude of
$\Sigma^{\prime\prime}$, by about 20\% (dashed line), but
approximately reproduces the shape of $\Sigma^{\prime\prime}$.
Hence we estimate that the average glue function extracted from optical
spectra should have the correct frequency dependence, but could be
overestimated by $\sim$20\% in intensity.

\begin{figure}[htop]
\hspace{-1cm}
\rotatebox{0}{\scalebox{.55}{\includegraphics{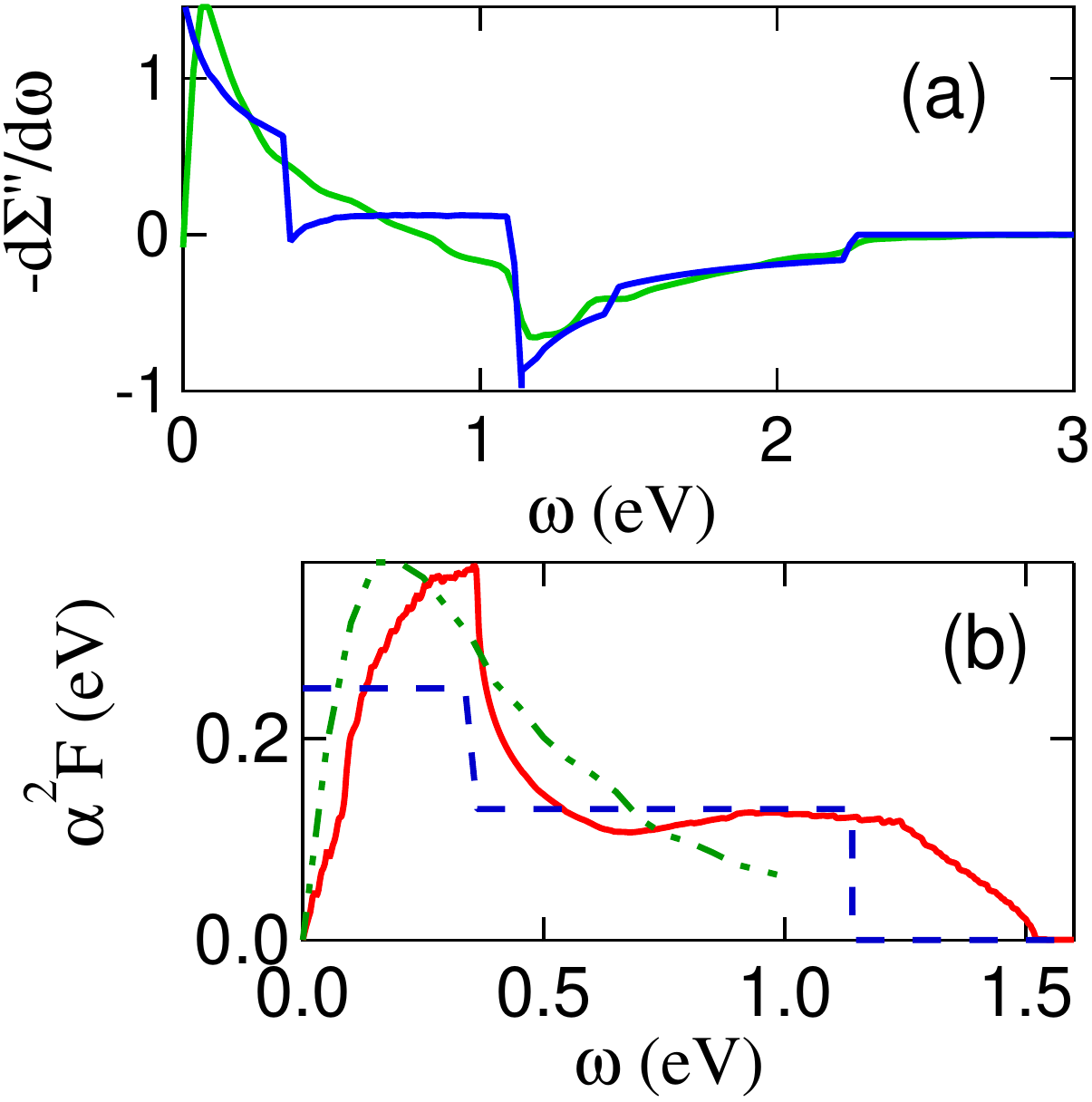}}}
\caption{(color online) (a)
$-\partial\Sigma^{\prime\prime}/\partial\omega$ (green line)
compared with fit based on Eq.~\protect\ref{eq:7} (blue line). (b)
Glue functions, comparing the calculated $\alpha^2F$ (blue dashed
line) with the input function from Fig.~\ref{fig:3}(a) (red line), and with an earlier optical glue function\cite{Casek} (green
dot-dot-dashed line).  }
\label{fig:3}
\end{figure}

Once the self energy is known, we can invert it to try to recover the susceptibility.  Here as a test case we take a derivative of $\Sigma$, Eq.~\ref{eq:4}, and fit the
$-\partial\Sigma^{\prime\prime}/\partial\omega$ data using a bar
graph representation for $\bar\chi^{\prime\prime}$.\cite{Marel} We
illustrate the calculation with only two bars, which works
reasonably well, but the generalization to many bars is
straightforward.  Thus, if $\alpha^2F (\omega )$ = $\alpha^2F_1$
for $\omega<\omega_1$, $\alpha^2F_2$ for
$\omega_1\le\omega<\omega_2$, and 0 for $\omega\ge\omega_2$, then
\begin{widetext}
\begin{equation}
-{1\over 2}{{\partial\Sigma^{\prime\prime}(\omega)}\over{\partial\omega}}=
\begin{cases}
\alpha^2F_1N(\omega) & \omega<\omega_1
\\
\alpha^2F_1[N(\omega)-N(\omega-\omega_1)]+\alpha^2F_2N(\omega-\omega_1) &
\omega_1\le\omega<\omega_2
\\
\alpha^2F_1[N(\omega)-N(\omega-\omega_1)]+\alpha^2F_2[N(\omega-\omega_1)
-N(\omega-\omega_2)]&\omega\ge\omega_2.
\end{cases}
\label{eq:7}
\end{equation}
\end{widetext}
Figure~\ref{fig:3}(a) shows a fit using this procedure, with $-\partial\Sigma^{\prime\prime}/\partial\omega$ taken from the
correct self-energy, blue curve in Fig.~\ref{fig:1}(b).  For a simple two-step glue function, the fit is remarkably good, and could be further improved by adding more steps.  Figure~\ref{fig:3}(b) shows that the resulting $\alpha^2F$ is a reasonable reproduction
of the input form.  Exact agreement is not expected, since the calculated $\alpha^2F$ is the scaled average used to generate the approximate self energy, red dashed curve in Fig.~\ref{fig:1}(b), whereas the extracted glue function is based in the exact self energy (calculated with anisotropic susceptibility).
The similarity of the two glue functions provides additional evidence that
the angle averaged formula captures the essential physics.
The two peaks in $\alpha^2F$ have a simple interpretation, the lower peak,
below 0.5~eV, represents the fluctuation spectrum of the coherent part of
the band, while the peak near 1~eV represents the fluctuations responsible
for opening the charge-transfer gap at lower doping.  A glimpse at
Fig.~\ref{fig:1}(a) reveals that these are heavily concentrated near $(\pi ,\pi )$.

Most previous analyses of the optical glue are consistent with the coherent part
of Fig.~\ref{fig:3}(b), displaying a peak near 0.3~eV.  For instance, the dot-dot-dashed
line in Fig.~\ref{fig:3}(b) shows the `continuum'
glue function extracted from optical experiments in Ref.~\onlinecite{Casek}.  However, they also find a
sharper peak at low energies, below $\sim$0.1~eV, not reproduced by the
present calculation.  We suggest that this peak may be associated either
with phonons or with superconducting or pseudogap effects [e.g., related to the
magnetic resonance peak], none of which are included in the present
normal state analysis.  We note that when superconductivity is included, our susceptibility calculations can
reproduce many features of the magnetic resonance phenomenon\cite{tanmoymagres}.

\subsection{The Weak Link: From $\Sigma_{op}$ to $\Sigma$}

Finally, we test how well the true self energy can be extracted from $\Sigma_{op}$ in Eq.~\ref{eq:0}.  First, we
expand Eq.~\ref{sigma} in the small $\omega$ limit, in which case $n_F(\omega^{\prime})-n_F(\omega^{\prime}+\omega)\simeq\omega
\delta(\omega^{\prime})$ at $T=0$, while $\omega+\Sigma^*(\omega^{\prime})
-\Sigma(\omega^{\prime}+\omega)=\omega\left(1-
\partial \Sigma^{\prime}(\omega)/\partial\omega\right) - 2i
\Sigma^{\prime\prime}(\omega)$, so that Eq.~\ref{sigma} becomes
\begin{equation}\label{eq:allen1}
\sigma(\omega) = \frac{i\omega_p^2}{4\pi}\frac{1}
{\omega-\omega\left(\partial\Sigma^{\prime}(\omega)/\partial\omega\right)-2i
\Sigma^{\prime\prime}(\omega)}.
\end{equation}
Comparing Eq.~\ref{eq:0} and Eq.~\ref{eq:allen1},
\begin{eqnarray}\label{opse}
\Sigma_{op}^{\prime}(\omega) &=&
\left(\partial\Sigma^{\prime}(\omega)/\partial
\omega\right)\omega /2 \\
\label{opse2}
\Sigma_{op}^{\prime\prime}(\omega) &=& \Sigma^{\prime\prime}(\omega).
\end{eqnarray}
In the special case where $\Sigma^{\prime}$ is quadratic in
$\omega$, $\Sigma_{op}=\Sigma$, the Method I result.  However, we generally find
$\Sigma^{\prime}\sim\omega$ at low frequencies, so
$\Sigma^{\prime}_{op}$ and $\Sigma^{\prime}$ differ by a factor of
2, and $\Sigma_{op}$ does not satisfy the Kramers-Kronig relation.

\begin{figure}[htop]
\rotatebox{0}{\scalebox{.65}{\includegraphics{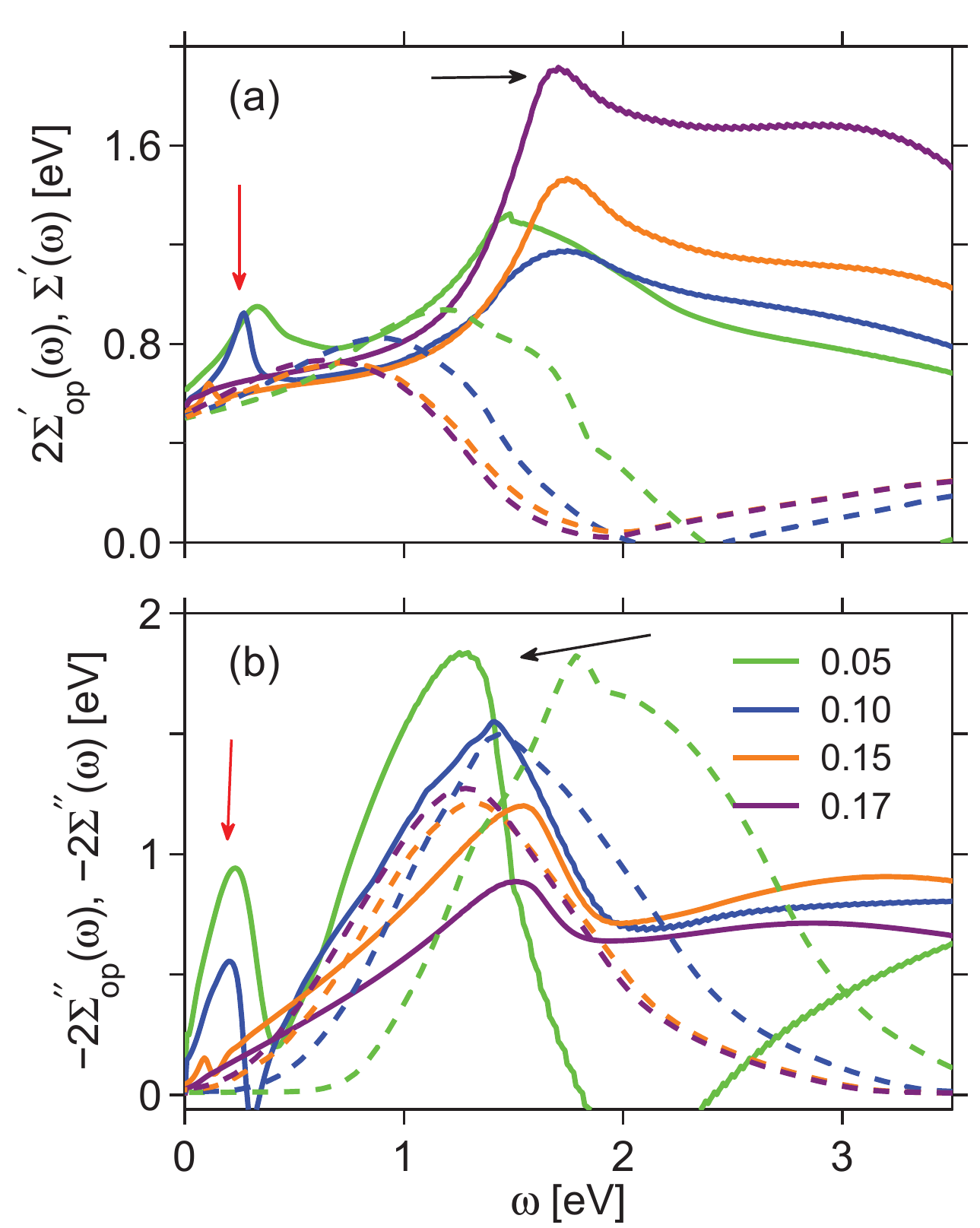}}}
\caption{(color online) The computed values of the optical self-energies
(solid lines) are compared with the corresponding quasiparticle self-energy
(dashed lines of same color) for a series of dopings for NCCO\cite{foot9,tanmoysw}. Both
results agree well in the low-energy, high-doping region.
}\label{fig:5}
\end{figure}

Despite this limitation, $\Sigma_{op}$ can be used to extract the
quasiparticle self-energy, at least in the low energy regime at
not-too-low doping.  Figure~\ref{fig:5} compares the quasiparticle self
energy $\Sigma$ of NCCO to $\Sigma_{op}$
calculated by first computing the optical
conductivity\cite{tanmoysw}, and then extracting $\Sigma_{op}$ via
Eq.~\ref{eq:0}. The optical conductivity is calculated from a standard
linear response theory in the presence of an antiferomagnetic
pseudogap and the quasiparticle self-energy $\Sigma$ corrections.
In accord with experimental results in NCCO, the optical spectra
show two distinct features, a mid-infrared feature originating
from the pseudogap order and the high-energy Mott gap feature
associated with the magnon scattering peak in
$\Sigma^{\prime\prime}$.\cite{tanmoysw}

Figure~\ref{fig:5} shows that Eqs.~\ref{opse} and~\ref{opse2}  hold in the low-energy region for the
paramagnetic phase.  However, for the underdoped samples
($x=0.05,0.10,0.15$) $\Sigma_{op}$ shows an additional kink at
low-energies coming from the pseudogap feature, not present in the
quasiparticle self-energy. [The calculated quasiparticle self-energy
includes the antiferromagnetic pseudogap, but this arises in the
off-diagonal term of a $2\times 2$ tensor, and is not captured in the
scalar approximation $\Sigma_{op}$.]  Thus, near optimal doping optical
studies should be able to determine $\Sigma"$ in the range up to
$\sim$1~eV, but the calculation breaks down in the underdoped regime.  For work at higher energies a more sophisticated approach is needed.  One possibility would be to use model self energies to reproduce the experimental spectra.
We have illustrated the comparison for Method I, but very similar results are found for Method II.

\section{Magnitude of Self Energy in ARPES and Optical Studies}

\begin{figure}[htop]
\rotatebox{0}{\scalebox{.65}{\includegraphics{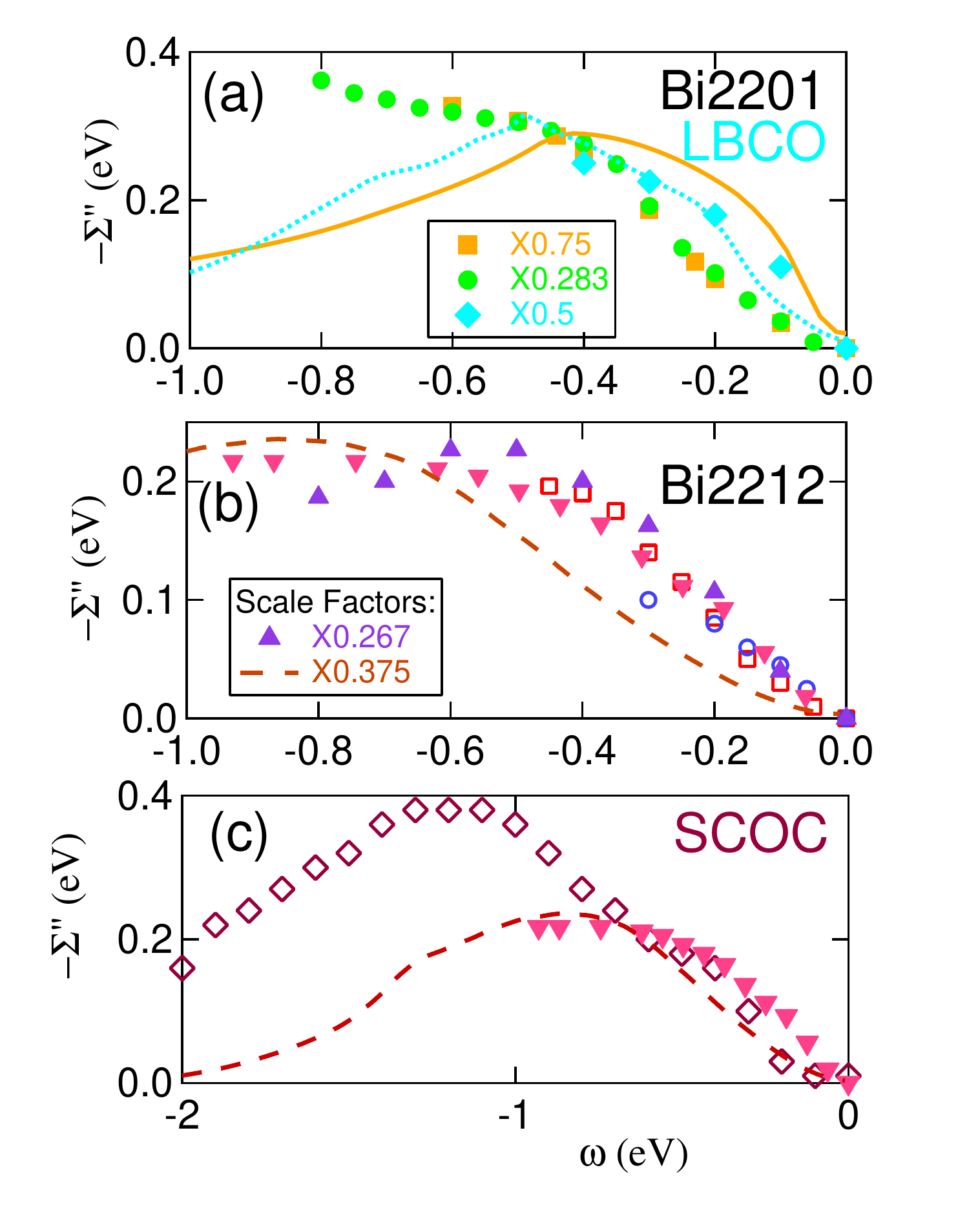}}}
\caption{(color online) Imaginary self energy $\Sigma^{\prime\prime}$
at $T=0$ vs energy $\omega$, comparing experimental and theoretical
results derived by several techniques.  In all cases, an impurity
contribution was approximately removed by subtracting off $\Sigma''(0)$.
Experimental points are ARPES data from: (a) LBCO (blue diamonds,
Ref.~\protect\onlinecite{Valla}); Bi2201 (gold squares,
Ref.~\protect\onlinecite{Non}, green circles,
Ref.~\protect\onlinecite{Feng}); (b) Bi2212 (violet triangles,
Ref.~\protect\onlinecite{Valla}, open red squares,
Ref.~\protect\onlinecite{Zhou}, open blue circles,
Ref.~\protect\onlinecite{Kordyuk}), and (c) Ca$_2$CuO$_2$Cl$_2$ (CCOC) (open
red-brown diamonds, Ref.~\protect\onlinecite{Ronning}).
Included in (b) and (c) are optical data from Bi2212 (inverted red
triangles, Ref.~\protect\onlinecite{vdMMZ}) (taken at $T=130K>T_c$ to avoid
complications associated with superconductivity). Theoretical curves are from:
LSCO (light blue dotted line), Ref.~\protect\onlinecite{tanmoyop},
Bi2201 (gold line), Ref.~\protect\onlinecite{water3},
Bi2212 (red dashed line), Ref.~\protect\onlinecite{waterfall}.
 Note that the magnitudes of several data sets have been rescaled.
} \label{fig:0}
\end{figure}

While the energy dependence of the self energy is readily extracted from optical or ARPES experiments, we find that there are subtle issues in normalizing the spectra.  ARPES probes the one-particle self energy, with full momentum dependence,
but only for the filled states.\cite{abfoot1,abarpes,abstm,abixs,abpositron}
Figure~\ref{fig:0} shows measured\cite{Valla,Non,Feng,Zhou,Kordyuk,Ronning} and calculated values\cite{foot8,foot9}
of the imaginary self energy $\Sigma"$
as a function of excitation energy $\omega$.  The data represent a number
of different cuprates at several dopings, but in all cases $\Sigma"
(\omega)$ has a similar shape.  There is a clear but relatively weak material
dependence, in good agreement with calculations.

However, the experiments fail to find a consistent {\it magnitude} of
$\Sigma"$, with values varying by a factor of four -- sometimes (as in the case of Bi2201, Fig.~\ref{fig:0}(a)) on
virtually the same material measured by two different groups.  This is
because $\Sigma"$ is not measured directly.  Instead, data were acquired
by measuring the momentum-space width $\Delta k$ of a spectral peak and
then multiplying by the `bare Fermi velocity', $v_{F0}$.
Unfortunately, the bare velocity is not a measured quantity, and a variety
of techniques have been utilized for estimating the value.
The largest
$\Sigma$s are found by assuming the bare and dressed dispersions do not
cross, hence drawing the bare dispersion as a straight line that either
lies below the ARPES dispersion or touches it at some high energy.
This assumption is equivalent, via Kramers-Kronig, to assuming that
$\Sigma"$ is monotonic in energy.  Neither of these features is consistent with theory -- see Ref.~\onlinecite{water3} and Fig.~\ref{fig:0} -- and moreover, most experiments find that the
ARPES linewidth narrows again at higher energies -- i.e., $\Sigma"$ should
have a peak in the region of the HEK.  Note that for a band of finite width it can be shown
analytically that $\Sigma''\rightarrow 0$ as $\omega\rightarrow\infty$.

Using a smaller $v_{F0}$ brings the
result into better agreement with theory.  The best choice for the bare
dispersion is probably the first principles LDA calculation.\cite{foot0,Kent}
However, it is also possible for experiments to find a too small value of $v_{F0}$.  This is
because in Bi2212 the self energy is large enough that the peak in $\Sigma"$ [the
high-energy kink] splits the dispersion into low and high energy branches, with a pseudogap in between.  ARPES experiments find the coherent band in Bi2212 is renormalized by a factor of $Z=0.5$ with coherent spectral weight extending to a band bottom at the
$\Gamma$-point near -0.5~eV,\cite{Zhou2,Dessau} whereas incoherent spectral weight extends
$\sim >1$~eV below the Fermi level.
If only this coherent branch is considered in extracting the self-energy,
the resulting $\Sigma"$ will be underestimated by the same factor $Z=0.5$,
consistent with the smallest values found in Fig.~\ref{fig:0}.

The Dresden group\cite{Kordyuk} attempted to extract both self energy and bare dispersion, treating the $\Gamma\rightarrow (\pi /2,\pi /2)$ (nodal) bare dispersion as a parabola, and adjusting the magnitude of the dispersion until the real and imaginary self energies satisfied Kramers-Kronig relations.  The resulting bare band can be parametrized by the energy of the band bottom at $\Gamma$, $E_0(\Gamma )=-0.9~eV$ for Bi-2212 at optimal doping.  In contrast, the corresopnding LDA result is -1.55~eV. This suggests either that LDA is not a good model for the bare bands or that the extracted self energies are too small by approximately the ratio of the $E_0(\Gamma )$'s, 0.6.  This is indeed close to the difference seen between their experiment and theory in Fig.~\ref{fig:0}.

We note that our calculations involve only electronic bosons, whereas experiments suggest that phonons may play a role in the temperature dependent broadening of the optical spectra, even in the undoped insulators\cite{phonons}.  This could explain some of the differences between calculated and experimental self energies.

 Figures~\ref{fig:0}(a,b) also display the optically derived $\Sigma"=-1/2\tau$
from Ref.~\onlinecite{vdMMZ}.  We see that its energy dependence is in
good agreement with theory, but its magnitude is smaller than theory by a
factor of 2, consistent with some ARPES evaluations.
In optical studies one can encounter a similar problem to those found in photoemission.
If $m^*(\omega )=m^*_1+m^*_2(\omega)$, then we can rewrite
\begin{equation}
\sigma(\omega) = \frac{i\omega_p^{*2}}{4\pi}\frac{1}{\omega (1+m_2^{**}/m)
 +i/\tau^*(\omega)},
\label{eq:0c}
\end{equation}
where $\omega_p^{*2}=Z\omega_p^{2}$, $m_2^{**}=Zm_2^*$, and
$1/\tau^*=Z/\tau$, with $Z=m/m_1^*$.
We note that if $\omega_p^*$ is used in Eq.~\ref{eq:0}, then the frequency dependence of
the extracted $\Sigma''$ is correctly given by
the measured $\sigma$, but its magnitude is too small by the factor$Z$.

\section{Discussion}


The present study finds that both low and high energy fluctuations couple strongly to
electronic excitations in the cuprates.  This has important implications for the origin of
superconductivity in these materials, and in particular is suggestive of two-component
$\alpha^2F$ models, with a strong peak
at low frequencies and a weak [electronic] peak at very high
frequencies.\cite{Carb1,Carb2} In Ref.\onlinecite{MRsc}, we
showed that in a hole-overdoped cuprate $x=0.3$, the glue functions below and above 0.3~eV made
comparable contributions to the supercondicting gap.  This is consistent with the predictions of Refs.~\onlinecite{PWA,glue}, but is  contradicted by another study which finds that low energy fluctuations in the vicinity of
the magnetic resonance peak can by themselves produce a 100K superconductor.\cite{Dahm}  The difference would seem to be that the latter study explored only the role of fluctuations near the resonance, whereas our full susceptibility calculation found an important role of ferromagnetic pairbreaking fluctuations, widely expected to be limiting $T_c$ on the overdoped side.\cite{TallStorey,Chakra}  As Cohen and Anderson\cite{CohAn} have noted, a key impediment to finding high $T_c$ superconductors is the emergence of competing phases.

\section{Conclusions}

In conclusion, our study provides a number of insights into attempts to derive self energies
and glue functions from experimental studies of the cuprates.  We find a surprising variability in
the magnitude of self energy reported in different studies.  We have identified possible sources and
recommend use of first principles dispersions in the analyses to minimize the problem.  With respect to
specifically optical studies, we find that the self energy is relatively momentum independent, so these
studies should be useful for extracting momentum averaged fluctuation spectra.  A possible weak link
is relating the optical spectrum to an underlying self energy, as the usual $\Sigma_{op}$ is found to
deviate from the true $\Sigma$ at energies above $\sim$1~eV.

Most importantly, we have shown that in the overdoped regime the optical spectra of the cuprates
should be described in a single band model for energies below $\sim$2.5~eV.  When this
is done, the resulting fluctuation spectrum or glue function displays substantial spectral weight in the high energy
region extending to $\sim$1.5~eV. Our study thus finds additional high
energy bosonic fluctuations in the optical spectra and reconciles a puzzling discrepancy in
this regard involving optical and other spectroscopies. In the conventional terminology of
optical studies, these contributions should also be considered as part of the optical `glue' function.

Of course, from optical studies there is no way of determining which of the observed fluctuations promote
d-wave superconductivity, which play no role in superconductivity, and which are actually pairbreaking.
Nevertheless, we must abandon the common perception that optical studies ``prove'' that only low frequency bosonic fluctuations are important for high-$T_c$ superconductivity -- particularly when those studies are restricted to energies below $\sim$1~eV.

This work is supported by the U.S.D.O.E contract DE-FG02-07ER46352
and benefited from the allocation of supercomputer time at NERSC
and Northeastern University's Advanced Scientific Computation
Center (ASCC).  This work was begun while RSM was on sabbatical at the University of Rome,
partially funded by the Marie Curie Grant PIIF-GA-2008-220790 SOQCS.  RSM acknowledges
stimulating conversations with M. Grilli.

\appendix
\section{Model Self-energy for Cuprates}


A recent series of calculations has led to a reassessment of the strength
of correlations in cuprates.  It has been found that, within a single
band Hubbard model, $U\le 8t$ is too small to satisfy the Brinkman-Rice
criterion for a Mott transition, and the magnetic phase in the cuprates
is closer to a conventional [Slater-type] antiferromagnet.\cite{TBPS,MGu}
Here $U$ is the Hubbard $U$-parameter and $t$ is the nearest neighbor
hopping.  With doping, spectral weight is transferred from the `upper
Hubbard band' to low energies too rapidly to be consistent with a
$U=\infty$ Hubbard or a $t-J$ model.\cite{EMS}  In contrast, intermediate coupling
models can describe this anomalous spectral weight
transfer,\cite{comanac,ASWT} and more generally provide a good description
of angle resolved photoemission spectroscopy (ARPES) and optical spectra
over a wide doping and energy range.\cite{water3,tanmoyop,DMFT1,DMFT2}  In
these models, a self energy is derived either from dynamic mean-field
theory calculations or from a modified GW procedure.  Here we describe the modified GW self energy calculation.


In the metallic phase at high doping, the quasiparticle-GW
(QP-GW) self-energy $\Sigma$ is given by a convolution over the
green function $G$ and the interaction $W\sim U^2\chi$ as
\cite{SWZ,vignale,waterfall,arpes,thesis},
\begin{eqnarray}\label{selfeng}
&&\tilde\Sigma({\bf k},\sigma,i\omega_n)=\frac{1}{2}U^2Z
\sum_{{\bf q},\sigma^{\prime}}^
{\prime}\eta_{\sigma,\sigma^{\prime}}
\int_{0}^{\infty}\frac{d\omega_p}{2\pi}\nonumber\\
&&\tilde G({\bf k}+{\bf q},\sigma^{\prime},i\omega_n,\omega_p)
\Gamma({\bf k},{\bf q},i\omega_n,\omega_p){\rm Im}[\tilde\chi_{\rm
RPA}^{\sigma\sigma^{\prime}}({\bf q},\omega_p)].\nonumber\\
\end{eqnarray}
where $\sigma$ is the spin index and $\eta_{\sigma,\sigma^{\prime}}$ is
3 for the spin and 1 for the charge modes.  Extensions to the
antiferromagnetic and/or superconducting phases are described in the
references.  In the QP-GW-scheme, $\Sigma$ is calculated self
consistently, with  $G$ and $W$ calculated from an approximate self-energy
$\Sigma^{t}_0(\omega)=\left(1-Z^{-1}\right)\omega$, where the
renormalization factor $Z$ is adjusted self-consistently to match
the coherent (low energy) part of the
self-energy.\cite{markiecharge,water3,arpes}  The vertex
correction $\Gamma$ in Eq.~\ref{selfeng} is taken as (Ward's identity)
$\Gamma=1/Z$. We take the dispersions directly from LDA calculations
($\xi_{\bf k}$), accurately fitted by a one band tight-binding
model\cite{markietb}, without any adjustment of the
resulting parameters.\cite{abfoot2,new1,new2}
In the overdoped regime a value for the screened Hubbard $U=1eV$
is used\cite{tanmoyop}.  In Eq.~\ref{selfeng} we use the RPA magnetic and
charge susceptibilities. Since the $k$-dependence of $\Sigma$ is weak\cite{water3},
we further simplify the calculation by assuming a $k$-independent
$\Sigma$, which we calculate at a representative point
$k=(\pi/2,\pi/2)$.

When the correct susceptibility is replaced by a
$k$-averaged version, the formula for the self energy simplifies.
The GW self energy can be written:
\begin{eqnarray}
\Sigma ({\bf q},\omega )&=&-{3\over 2}U^2\sum_{\bf
k}\int_{-\infty}^{\infty}\chi''({\bf k},\Omega )d\Omega
\int_{-\infty}^{\infty}A({\bf k+q},\epsilon )d\epsilon
\nonumber \\
&&\times\Bigl[{n_B(\Omega )+n_F(\epsilon )\over \omega+\Omega
-\epsilon}+{n_B(\Omega )+1-n_F(\epsilon )\over \omega-\Omega
-\epsilon}
\Bigr],
\label{eq:3a}
\end{eqnarray}
where $A({\bf k},\epsilon )$ is the electronic spectral function, $n_F$
[$n_B$] is the Fermi [Bose] function,
and $\chi^{\prime\prime}$ is the imaginary part of an appropriate
susceptibility.\cite{Mah,water3} Then
\begin{eqnarray}
\Sigma^{\prime\prime}({\bf q},\omega )&=&-{3\over 2}U^2\sum_{\bf
k}\int_{-\infty}^{\infty}\chi''({\bf k},\Omega )d\Omega
\nonumber \\
&&\times\Bigl[A({\bf k+q},\omega+\Omega )(n_B(\Omega )+n_F(\omega+\Omega ))
\nonumber \\
&&+A({\bf k+q},\omega-\Omega )(n_B(\Omega )+1-n_F(\omega-\Omega ))
\Bigr].
\label{eq:3b}
\end{eqnarray}

When $\chi"$ is replaced by $\bar\chi"$ in Eq.~\ref{eq:3b}, the ${\bf k}$-sum
reduces to $\sum_{\bf k}A({\bf k},\epsilon )=N(\epsilon )$ and Eq.~\ref{eq:3b}
becomes\cite{SharC,Hwang}
\begin{eqnarray}
\Sigma^{\prime\prime}(\omega )&=&-{1\over
2}\int_{-\infty}^{\infty}\frac{\alpha^2F(\Omega )}{N_{av}}d\Omega
\nonumber \\
&&\times\Bigl[N(\omega+\Omega )\bigl({\rm coth}({\Omega\over
2T})-{\rm tanh}({\Omega+\omega\over 2T})\bigr)
\nonumber \\
&&+N(\omega-\Omega )\bigl({\rm coth}({\Omega\over 2T})-{\rm tanh}({\Omega-\omega\over
2T})\bigr)\Bigr].
\label{eq:5}
\end{eqnarray}
At $T=0$ and $\omega >0$, this becomes Eq.~\ref{eq:4}, while
for $\omega <0$, the only changes are that the upper limit of the
integral is $|\omega |$ and the argument of $N$ is $\Omega
-|\omega |$ which is $<0$.


\begin{thebibliography}{99}
\bibitem{resy2}H. Woo, P. Dai,  S.M. Hayden, H.A. Mook, T. Dahm, D.J. Scalapino, T.G. Perring, and F. Dogan, Nature Physics {\bf 2}, 600 (2006).
%
\bibitem{PWA}P.W. Anderson, Science {\bf 316}, 1705 (2007).
%
\bibitem{glue}T.A. Maier, D. Poilblanc, and D.J. Scalapino, Phys. Rev. Lett. {\bf 100}, 237001 (2008).
%
\bibitem{MRsc} R. S. Markiewicz and A. Bansil, Phys. Rev. B {\bf 78},
134513 (2008).
%
\bibitem{Emery}V. J. Emery, Phys. Rev. Lett. {\bf 58}, 2794 (1987); V. J. Emery and G. Reiter, Phys. Rev. B{\bf 38}, 4547 (1988).
%
\bibitem{SchaCar}E. Schachinger and J.P. Carbotte, Phys. Rev. B{\bf 62}, 9054 (2000).
%
\bibitem{DoHoTV}S.V. Dordevic, C.C. Homes, J.J. Tu, T. Valla, M. Strongin, P.D. Johnson, G.D. Gu, and D.N. Basov, Phys. Rev. B{\bf 71}, 104529 (2005).
%
\bibitem{Casek}P. C\'asek, C. Bernhard, J. Huml�{\v c}ek, and D. Munzar, Phys. Rev. B{\bf 72}, 134526 (2005).
%
\bibitem{NoChu}M.R. Norman and A.V. Chubukov, Phys. Rev. B{ 73}, 140501
(2006).
\bibitem{Marel}E. van Heumen, E. Muhlethaler, A.B. Kuzmenko, H. Eisaki, W. Meevasana, M. Greven, and D. van der Marel, Phys. Rev. B{\bf 79}, 184512 (2009).
%
\bibitem{GDM}C. Giannetti, F. Cilento, S. Dal Conte, G. Coslovich, G. Ferrini, H. Molegraaf, M. Raichle, R. Liang, H. Eisaki, M. Greven, A. Damascelli, D. van
 der Marel, and F. Parmigiani, Nature Communications {\bf 2}, 353 (2011).
%
\bibitem{Seam}Jinho Lee, K. Fujita, K. McElroy, J.A. Slezak, M. Wang, Y. Aiura, H. Bando, M. Ishikado, T. Masui, J.-X. Zhu, A.V. Balatsky, H. Eisaki, S. Uchida and J.C. Davis, Nature (London) {\bf 442}, 546 (2006).
%
\bibitem{Lanz}G.-H. Gweon, T. Sasagawa, S.Y. Zhou, J. Graf, H. Takagi, D.-H. Lee, and A. Lanzara, Nature (London) {\bf 430}, 187 (2004).
%
\bibitem{DanD}H. Iwasawa, J.F. Douglas, K. Sato, T. Masui, Y. Yoshida, Z. Sun, H. Eisaki, H. Bando, A. Ino, M. Arita, K. Shimada, H. Namatame, M. Taniguchi, S. Tajima, S. Uchida, T. Saitoh, D.S. Dessau, and Y. Aiura, Phys. Rev. Lett. {\bf 101}, 157005 (2008); K. Sato, H. Iwasawa, N.C. Plumb, T. Masui, Y. Yoshida, H. Eisaki, H. Bando, A. Ino, M. Arita, K. Shimada, H. Namatame, M. Taniguchi, S. Tajima, Y. Nishihara, D.S. Dessau, and Y. Aiura, Phys. Rev. B{\bf 80}, 212501 (2009).
%
\bibitem{RonK}F. Ronning, K.M. Shen, N.P. Armitage, A. Damascelli, D.H. Lu, Z.-X. Shen, L.L. Miller, and C. Kim, Phys. Rev. B{\bf 71}, 094518 (2005).
%
\bibitem{Ale}J. Graf, G.-H. Gweon, K. McElroy, S.Y. Zhou, C. Jozwiak, E. Rotenberg, A. Bill, T. Sasagawa, H. Eisaki, S. Uchida, H. Takagi, D.-H. Lee, and A. Lanzara, Phys. Rev. Lett. {\bf 98}, 067004 (2007).
%
\bibitem{Non}W. Meevasana, X.J. Zhou, S. Sahrakorpi, W.S. Lee, W.L. Yang, K. Tanaka, N. Mannella, T. Yoshida, D.H. Lu, Y.L. Chen, R.H. He, Hsin Lin,
S. Komiya, Y. Ando, F. Zhou, W.X. Ti, J.W. Xiong, Z.X. Zhao, T. Sasagawa, T. Kakeshita, K. Fujita, S. Uchida, H. Eisaki, A. Fujimori, Z. Hussain, R.S. Markiewicz, A. Bansil, N. Nagaosa, J. Zaanen, T.P. Devereaux, and Z.-X. Shen, Phys. Rev. B.{\bf 75}, 174506 (2007).
%
\bibitem{Valla}T. Valla, T.E. Kidd, W.-G. Yin, G.D. Gu, P.D. Johnson, Z.-H. Pan, and A.V. Fedorov, Phys. Rev. Lett. {\bf 98}, 167003 (2007).
%
\bibitem{Feng}B.P. Xie, K. Yang, D.W. Shen, J.F. Zhao, H.W. Ou, J. Wei, S.Y. Gu, M. Arita, S. Qiao, H. Namatame, M. Taniguchi, N. Kaneko, H. Eisaki, Z.Q. Yang, and D.L. Feng, Phys. Rev. Lett. {\bf 98}, 147001 (2007). 
%
\bibitem{Zhou}J.M. Bok, J.H. Yun, H.-Y. Choi, W. Zhang, X.J. Zhou, and C.M. Varma, Phys. Rev. B{\bf 81}, 174516 (2010).
%
\bibitem{Kordyuk}A.A. Kordyuk, S.V. Borisenko, V.B. Zabolotnyy, J. Geck, M. Knupfer, J. Fink, B. B\"uchner, C.T. Lin, B. Keimer, H. Berger, A.V. Pan, S. Komiya, and Y. Ando, Phys. Rev. Lett. {\bf 97}, 017002 (2006).
%
\bibitem{Ronning}C. Kim, S.R. Park, C.S. Leem, D.J. Song, H.U. Jin, H.-D. Kim, F. Ronning, and C. Kim, Phys. Rev. B{\bf 76}, 104505 (2007).
%
\bibitem{water3}R.S. Markiewicz, S. Sahrakorpi, and A. Bansil, Phys. Rev. B {\bf 76 }, 174514 (2007).
%
\bibitem{waterfall} Susmita Basak, Tanmoy Das, Hsin Lin, J. Nieminen, M. Lindroos, R. S. Markiewicz, and A. Bansil, Phys. Rev. B {\bf 80}, 214520 (2009).
%
\bibitem{foot1}Notably, optical techniques cannot access the full $\alpha^2F({\bf q},\omega )$ needed for the symmetry averaged $d$-wave
glue; see D.J. Scalapino, E. Loh, Jr., and J.E. Hirsch, Phys. Rev. B{\bf 34}, 8190 (1986); {\bf 35}, 6694 (1987) and Ref.~\onlinecite{MRsc}.
%
\bibitem{foot4}For example, Ref.~\onlinecite{GDM} presents optical spectra for Bi2212 extending to $\sim$6~eV, with a peak associated with the residual charge-transfer gap identified near 2.72~eV.  While features below this energy are argued not to involve $d-d$ transitions, the spectrum is arbitrarily divided into an `intraband' part below 1.25~eV and an `interband' part at higher energies, and only the former is assumed to contribute to the glue function.
\bibitem{tanmoyop} Tanmoy Das,  R.S. Markiewicz, and A. Bansil, arXiv:0807.4257, Phys. Rev. B {\bf 81}, 174504 (2010).
%
\bibitem{comanac} A. Comanac, L. de Medici, M. Capone, and A. J. Millis, Nature Physics, {\bf 4} 287 (2008); L. de' Medici, X. Wang, M. Capone, and A.J. Millis, Phys. Rev. B {\bf 80}, 054501 (2009).
%
\bibitem{DMFT1}N. Lin, E. Gull, and A. J. Millis, Phys. Rev. B{\bf 82}, 045104 (2010), and references therein.
%
\bibitem{DMFT2}C. Weber, K. Haule, and G. Kotliar, Nat. Phys. {\bf  6}, 574 (2010).
%
\bibitem{Little}M.J. Holcomb, J.P. Collman, and W.A. Little, Phys. Rev. Lett. {\bf 73}, 2360 (1994); H.J.A. Molegraaf, C. Presura, D. van der Marel, P.H. Kes, and M. Li, Science {\bf 295}, 2239 (2002).
%
\bibitem{Alle}P.B. Allen, arXiv: 0407777.
%
\bibitem{hwangnature}J. Hwang, T. Timusk, and G.D. Gu, Nature {\bf 427}, 714 (2004).
%
\bibitem{hwangprl} J. Hwang, E.J. Nicol, T. Timusk, A. Knigavko, and J.P. Carbotte, Phys. Rev. Lett. {\bf 98}, 207002 (2007).
%
\bibitem{PBA2}P.B. Allen, Phys. Rev. B{\bf 3}, 305 (1971).
%
\bibitem{Mars}F. Marsiglio, T. Startseva, and J.P. Carbotte, Phys. Lett. A {\bf 245}, 172 (1998); F. Marsiglio, J. Supercond. {\bf 12}, 163 (1999).
%
\bibitem{Shulga}S.V. Shulga, ``Material Science, Fundamental Properties and Future Electronic Applications of High-Tc Superconductors", (Kluwer Academic Publishers, Dortrecht, 2001), pp. 323-360 arXiv:cond-mat/0101243.
%
\bibitem{HEscale}J. Hwang, E.J. Nicol, T. Timusk, A. Knigavko, and J.P. Carbotte, Phys. Rev. Lett. {\bf 98}, 207002 (2007).
%
\bibitem{Hedin}L. Hedin, J. Phys. Condens. Matter {\bf 11}, R489 (1999).
%
\bibitem{foot6}This formula was generalized to finite temperatures by J. Hwang, T. Timusk, E. Schachinger, and J.P. Carbotte, Phys. Rev. B{\bf 75}, 144508 (2007).  Notably, however, they still assume a constant density of states, and hence overlook the correction discussed here.
%
\bibitem{foot2} A similar problem arises when the spectrum has a gap, see  Ref.~\onlinecite{DoHoTV}.
%
\bibitem{Maksimov}E.G. Maksimov, Physics - Uspekhi {\bf 43}, 965 (2000).
\bibitem{QT}M.A. Quijada, D.B. Tanner, R.J. Kelley, M. Onellion, H.
Berger, and G. Margaritondo, Phys. Rev. B{\bf 60}, 14917 (1999).
\bibitem{footC1}This agreement justifies the assumed background subtraction in Ref.~\onlinecite{tanmoyop}, and further suggests an interpretation of this background in terms of interband transitions to a $d_{z^2}$-like band, which is believed to be much closer to the Fermi level in LSCO than in other cuprates\cite{Aoki}. 
%
\bibitem{Aoki}H. Sakakibara, H. Usui, K. Kuroki, R. Arita, and H. Aoki, arXiv:1003.1770.
%
\bibitem{Uchida}S. Uchida, T. Ido, H. Takagi, T. Arima, Y. Tokura, and S. Tajima, Phys. Rev. B{\bf 43}, 7942 (1991).
%
\bibitem{footy}We comment briefly on the deviation between theory and experiment in Fig.~1.  First, we note that this was part of a comprehensive fit over the full doping range in Ref.~\onlinecite{tanmoyop}, and no effort was made to adjust the parameters here.  Second we note that the problem is an underestimate of the strength of the residual charge transfer band, as shown by the thin blue dashed line in Fig.~1(a). This weakness leads to the fast falloff of $\epsilon$ at higher frequencies, Fig.~1(c).  Since the zero crossing of $\epsilon$ fixes the plasmon peak, this leads to the peak in (b) being too sharp and at too low an energy.  There are three possible sources of discrepancy.  First, it could be that some of the parameters used in
 Ref.~\onlinecite{tanmoyop} have a weak doping dependence, which was neglected.  Secondly, the QP-GW method is an approximation, which is designed to accurately capture the coherent part of the self energy, but at the cost of underestimating the high-energy incoherent part of the self energy.  Thirdly, this calculation approximated the dielectric constant in the Hubbard model as $1+U\chi$, whereas we know that including long-range Coulomb interactions is important for the high-energy effects of the charge susceptibility.\cite{markiecharge}
 %
\bibitem{markiecharge} R.S. Markiewicz, and A. Bansil, Phys. Rev. B {\bf 75}, 020508 (2007).
%
\bibitem{footC2}The feature in Bi2212 above 3.5~eV was suggested to be related to BiO-layer interband transitions,\cite{QT} and we confirm that it is not present in the one-band Hubbard model.
%
\bibitem{vdMMZ}D. van der Marel, H.J.A. Molegraaf, J. Zaanen, Z. Nussinov, F. Carbone, A. Damascelli, H. Eisaki, M. Greven, P.H. Kes, and M. Li, Nature {\bf 425}, 271 (2003).
%
\bibitem{MarB2} R.S. Markiewicz, M.Z. Hasan, and A. Bansil, Phys. Rev. B {\bf 77}, 094518 (2008).
%
\bibitem{ChaMu}J. Chaloupka and D. Munzar, Phys. Ref. B{\bf 76}, 214502 (2007).
%
\bibitem{foot3}In these calculations, we work near optimal hole doping using parameters appropriate for NCCO, but most cuprates have similar susceptibilities.\cite{MGu2}
%
\bibitem{MGu2}R.S. Markiewicz, J. Lorenzana, G. Seibold, and A. Bansil, Phys. Rev. B{\bf 81}, 014509 (2010).
%
\bibitem{foot9}Since the momentum dependence of $\Sigma$ is weak, we take the result for $q=(\pi /2,\pi /2)$ as representative.
\bibitem{tanmoymagres}T. Das, R.S. Markiewicz, and A. Bansil, unpublished; A. Bansil, Susmita Basak, Tanmoy Das, Hsin Lin, M. Lindroos, J. Nieminen, Ilpo Suominen, and R.S. Markiewicz, J. Phys. Chem. Solids {\bf 71}, 341 (2011).
%
\bibitem{tanmoysw} Tanmoy Das, R.S. Markiewicz and A. Bansil, Phys. Rev. B {\bf 81}, 184515 (2010).
%
\bibitem{abfoot1} We are not taking into account matrix element effects, which in general can be important in ARPES\cite{abarpes}, STM\cite{abstm} and other highly resolved spectroscopies\cite{abixs,abpositron}.
%
\bibitem{abarpes} S. Sahrakorpi, M. Lindroos, R. S. Markiewicz, and A. Bansil, Phys. Rev. Lett. {\bf 95}, 157601 (2005); J. C. Campuzano, L.C. Smedskjaer, R. Benedek, G. Jennings, and A. Bansil, Phys. Rev. B {\bf 43}, 2788 (1991); A. Bansil, M. Lindroos, S. Sahrakorpi, and R.S. Markiewicz, Phys. Rev. B {\bf 71}, 012503 (2005).
%
\bibitem{abstm}J. Nieminen, H. Lin, R. S. Markiewicz, and A. Bansil, {Phys. Rev. Lett.} {\bf 102}, 037001 (2009).
%
\bibitem{abixs} R. S. Markiewicz and A. Bansil, Phys. Rev. Lett. {\bf 96}, 107005 (2006); S. Huotari, K. Hamalinen, S. Manninen, S. Kaprzyk, A. Bansil, W. Caliebe, T. Buslaps, V. Honkimaki, and P. Suortti, Phys. Rev. B {\bf 62}, 7956 (2000); P. E. Mijnarends and A. Bansil, Phys. Rev. B {\bf 13}, 2381 (1976).
%
\bibitem{abpositron} P. E. Mijnarends, A. C. Kruseman, A. van Veen, H. Schut and A. Bansil, J. Physics: Condens. Matter 10, 10383(1998); J. Mader, S. Berko, H. Krakauer and A. Bansil, Phys. Rev. Lett. {\bf 37}, 1232(1976); L. C. Smedskjaer, A. Bansil, U. Welp, Y. Fang, and K. G. Bailey, Physica C {\bf 192}, 259(1992).
%
\bibitem{foot8}We calculate the self energy due to magnetic fluctuations only, since these dominate within $\sim$1~eV of the Fermi level.
%
\bibitem{foot0}Care must be taken in extracting the bare $d_{x^2-y^2}$ band -- far below the Fermi level it anticrosses with several other bands, so the bare
dispersion must be chosen to follow the $d_{x^2-y^2}$ spectral weight.\cite{Kent}
%
\bibitem{Kent}P.R.C. Kent, T. Saha-Dasgupta, O. Jepsen, O.K. Andersen, A. Macridin, T.A. Maier, M. Jarrell, and T.C. Schulthess, Phys. Rev. B{\bf 78} 035132 (2008).
%
\bibitem{Zhou2}W. Zhang, G. Liu, J. Meng, L. Zhao, H. Liu, X. Dong, W. Lu, J.S. Wen, Z.J. Xu, G.D. Gu, T. Sasagawa, G. Wang, Y. Zhu, H. Zhang, Y. Zhou, X. Wang, Z. Zhao, C. Chen, Z. Xu, and X.J. Zhou, Phys. Rev. Lett. {\bf 101}, 017002(2008).
%
\bibitem{Dessau}Q. Wang, Z. Sun, E. Rotenberg, H. Berger, H. Eisaki, Y. Aiura, and D.S. Dessau, arXiv:0910.2787.
%
\bibitem{phonons}R. L\"ovenich, A.B. Schumacher, J.S. Dodge, D.S. Chemla, and L.L. Miller,  Phys. Rev. B{\bf 63}, 235104 (2001).
%
\bibitem{Carb1}F. Marsiglio and J.P. Carbotte, Phys. Rev. B{\bf 36}, 3937 (1987).
%
\bibitem{Carb2}J.P. Carbotte, Rev. Mod. Phys. {\bf 62}, 1027 (1990).
%
\bibitem{Dahm}T. Dahm, V. Hinkov, S.V. Borisenko, A.A. Kordyuk, V.B. Zabolotnyy, J. Fink, B. Bchner, D.J. Scalapino, W. Hanke, and B. Keimer, Nature Physics {\bf 5}, 217 (2009).
\bibitem{TallStorey}J.G. Storey, J.L. Tallon, and G.V.M. Williams, Phys. Rev. B{\bf 76}, 174522 (2007).
%
\bibitem{Chakra}A. Kopp, A. Ghosal, and S. Chakravarty,  Proc. Natl. Acad. Sci. USA {\bf 104}, 6123 (2007).
%
\bibitem{CohAn}M.L. Cohen and P.W. Anderson, in {\it Superconductivity in d- and f-Band Metals}, ed. by D.H. Douglass (Am. Inst. Phys., N.Y., 1972), p. 17.
%
\bibitem{TBPS}L.F. Tocchio, F. Becca, A. Parola, and S. Sorella, Phys. Rev. B{\bf 78}, 041101(R) (2008); F. Becca, L.F. Tocchio, and S. Sorella, J. Phys.: COnf. Ser. {\bf 145}, 012016 (2009).
%
\bibitem{MGu}R.S. Markiewicz, J. Lorenzana, and G. Seibold, Phys. Rev. B{\bf 81}, 014510 (2010).
%
\bibitem{EMS} H. Eskes, M.B. Meinders, and G.A. Sawatzky, Phys. Rev. Lett. {\bf 67}, 1035 (1991).
%
\bibitem{ASWT}R.S. Markiewicz, Tanmoy Das, and A. Bansil, Phys. Rev. B{\bf 82}, 224501 (2010).
%
\bibitem{SWZ} J.R. Schrieffer, X.G. Wen, and S.C. Zhang, Phys. Rev. B {\bf 39}, 11663 (1989).
%
\bibitem{vignale} G. Vignale and M. R. Hedayati, Phys. Rev. B {\bf 42}, 786 (1990).
%
\bibitem{arpes} R.S. Markiewicz, T. Das, S. Basak, and A. Bansil, J. Elect. Spect. Rel. Phenom. {\bf 181}, 23 (2010).
%
\bibitem{thesis} Tanmoy Das, PhD Thesis, "A model of coexistence of antiferromagnetism and superconductivity in electron- and
hole-doped cuprates." Northeastern Univ., 2009 (unpublished).
%
\bibitem{markietb} R.S. Markiewicz, S. Sahrakorpi, M. Lindroos, Hsin Lin, and A. Bansil, Phys. Rev. B {\bf 72}, 054519 (2005).
%
\bibitem{abfoot2} The band structure is taken to be doping independent in the spirit of the rigid band model\cite{new1}], which is expected to be a good approximation for doping away from the CuO$_2$ planes. It will be interesting to examine doping effects via first principles approaches\cite{new2}.
%
\bibitem{new1} A. Bansil, Phys. Rev. B20, 4025(1979); A. Bansil, Phys. Rev. B20, 4035(1979); R. Prasad and A. Bansil, Phys. Rev. B21, 496 (1980); H. Asonen, M. Lindroos, M. Pessa, R. Prasad, R.S. Rao, and A. Bansil, Phys. Rev. B {\bf 25},
7075 (1982).
%
\bibitem{new2} S.N. Khanna, A.K. Ibrahim, S.W. McKnight, and A. Bansil, Solid State Commun. {\bf 55}, 223 (1985); L. Huisman, D. Nicholson, L. Schwartz and A. Bansil, Phys. Rev. B 24, 1824 (1981); L. Schwartz and A. Bansil, Phys. Rev. B {\bf 10}, 3261(1974).
%
\bibitem{Mah}G.D. Mahan, ``Many-Particle Physics'' (2d Ed.) (Plenum, New York, 1990).
%
\bibitem{SharC}S.G. Sharapov and J.P. Carbotte, Phys. Rev. B{\bf 72}, 134506 (2005).
%
\bibitem{Hwang}J. Hwang, Phys. Rev. B{\bf 83}, 014507 (2011).
%
\end{thebibliography}
\end{document}